\documentclass[aps,preprint,amsmath,amssymb]{revtex4}
\usepackage{graphicx}
\begin{document}

\title{Search for the electromagnetic moments of the $\tau$ lepton in photon-photon collisions
at the LHeC and the FCC-he}

\author{M. Koksal}
\email[]{mkoksal@cumhuriyet.edu.tr} \affiliation{Department of
Optical Engineering, Cumhuriyet University, 58140, Sivas, Turkey}

\begin{abstract}
We examine the potential of the process $ep \rightarrow e\gamma^{*} \gamma^{*} p\rightarrow e\tau \bar{\tau}p$  at the Large Hadron Electron Collider (LHeC) and the Future Circular Hadron Electron Collider (FCC-he) to examine non-standard $\tau\bar{\tau}\gamma$ coupling in a model independent way by means of the effective Lagrangian approach. We perform pure leptonic and semileptonic decays for $\tau^{-}\tau^{+}$ production in the final state. Furthermore, we use $L = 10, 50, 100, 300, 500, 1000$ fb$^{-1}$ with at $\sqrt{s}=1.30, 1.98, 7.07, 10$ TeV and we consider systematic uncertainties of $\delta_{sys}=0,5,10 \%$. The best sensitivity bounds obtained from the process $ep \rightarrow e\gamma^{*} \gamma^{*} p\rightarrow e\tau \bar{\tau}p$ on the anomalous couplings are $-0.0025<\tilde{a}_{\tau}<0.0009$ and $|\tilde{{d}_{\tau}}|<8.85 \times 10^{-18}\, e\,cm$, respectively. Therefore, our results show that the process $ep \rightarrow e\gamma^{*} \gamma^{*} p\rightarrow e\tau \bar{\tau}p$ at the LHeC and FCC-he are a very good prospect for probing the anomalous magnetic and electric dipole moments of the $\tau$ lepton at $\gamma^{*} \gamma^{*}$ mode of the future $ep$ collider.
\end{abstract}

\maketitle

\section{Introduction}

The magnetic dipole moment of the electron which is responsible for the interaction with the
magnetic field in the Born approximation is given as follows

\begin{eqnarray}
\vec{\mu}=g \frac{\mu_{B}}{\hbar}\vec{s}.
\end{eqnarray}
Here, $g$ is the Lande g-factor or gyromagnetic factor, $\mu_{B}$ is the Bohr magneton and $\vec{s}$ represents the spin of the electron. For the electron, the value of $g$ in the Dirac equation is $2$. It is traditional to point out the deviation of $g$ from 2 in terms of the value of the so-called anomalous
magnetic moment. The anomalous magnetic moment of the electron is a dimensionless quantity and is described by

\begin{eqnarray}
a_{e}=\frac{(g-2)}{2}.
\end{eqnarray}
The $a_{e}$ without anomalous and radiative corrections is equal to $0$. Besides, it was firstly found from Quantum Electrodynamics (QED) using radiative corrections by Schwinger as $a_{e}=\frac{\alpha}{2\pi}$ \cite{1}.

The accuracy of the $a_{e}$ has been studied so far in many works. These works have provided the most precise determination of fine-structure constant $\alpha_{QED}$, since $a_{e}$ is quite senseless to the strong and weak interactions. However, the $a_{\mu}$ anomalous magnetic moment of the muon enables testing the Standard Model (SM) and investigating alternative theories to the SM. Especially, the $a_{\mu}$ is more sensitive to new physics beyond the SM by a factor of $(\frac{m_{\mu}}{m_{e}})^{2}\sim4 \times 10^{4}$ than to the case of the $a_{e}$. Similarly, due to the large mass of the $\tau$ lepton, a precise measurement of the $a_{\tau}$ anomalous magnetic moment provides an excellent opportunity to reveal the effects of the new physics beyond the SM.

The $a_{e}$ and $a_{\mu}$ have been examined with high sensitivity through spin precession experiment. On the other hand, spin precession experiment is not appropriate to investigate the $a_{\tau}$ anomalous magnetic moment because of the relatively short lifetime $2.906 \times 10^{-13}$ s of the $\tau$ lepton \cite{tau}. Instead of those experiments, we focus on highly precise measurements by comparing the measured cross section with the SM cross section in colliders with high center-of-mass energies in $\tau^{-}\tau^{+}$ production processes.

The SM contribution for the $a_{\tau}$ anomalous magnetic moment is obtained by the sum of QED, electroweak and hadronic
terms. The theoretical contribution from the QED to the $a_{\tau}$ anomalous magnetic moment up to three loops is calculated as \cite{2,3,pas}

\begin{eqnarray}
a_{\tau}^{QED}=117 324 \times 10^{-8}.
\end{eqnarray}

In addition, the sum of the one and two loop electroweak effects is given by

\begin{eqnarray}
a_{\tau}^{EW}=47.4 \times 10^{-8}.
\end{eqnarray}

The hadronic contribution to the $a_{\tau}$ anomalous magnetic
moment arising from QED diagrams including hadrons is

\begin{eqnarray}
a_{\tau}^{HAD}=350.1 \times 10^{-8}.
\end{eqnarray}

By collecting all these additives, we obtain the $a_{\tau}$ anomalous magnetic moment

\begin{eqnarray}
a_{\tau}=a_{\tau}^{QED}+a_{\tau}^{EW}+a_{\tau}^{HAD}= 117 721 \times 10^{-8}.
\end{eqnarray}

The SM involves three sources of CP violation. One of them appears by complex couplings in the Cabibbo-Kobayashi-Maskawa (CKM) matrix of the quark sector  \cite{8}. In the SM, neutrinos are massless. With correction of the SM to contain neutrino masses, CP violation can occur in the mixing of leptons. The last source of this phenomenology is possible in flavor conserving strong interaction processes. Besides, the experimental upper
limit on the neutron electric dipole moment indicates that this ($\theta_{QCD}/16\pi^{2})F_{\mu\nu}F\tilde{^{\mu\nu}}$ term in the
SM Lagrangian is at best tiny, $\theta_{QCD}\ll 10^{-9}$. This is generally known as the strong CP problem. As mentioned above, although there is CP violation in the SM, it is not enough to explain for the observed baryon asymmetry of the universe given the limits on baryon number violation. It is clear that there must be CP violation beyond the SM.

The electric dipole moment of the $\tau$ lepton allows a direct investigation of CP violation \cite{9,10}, a property of the SM and new physics beyond the SM. However, CP violation in the quark sector induces a small electric dipole moment of the $\tau$ lepton. One has to go at least to three loop level to create a non-zero contribution. It’s crude estimate gives as follows \cite{111}

\begin{eqnarray}
|d_{\tau}|\leq 10^{-34}\, e\,cm.
\end{eqnarray}

Thus, the electric dipole moment of the $\tau$ lepton is undetectably small with the contributions arising from the SM. Besides, the electric dipole moment of this lepton may cause detectable size due to interactions arising from the new physics beyond the SM such as leptoquarks \cite{11,12}, supersymmetry \cite{13,yaman}, left-right symmetric models \cite{14,bok} and more Higgs multiplets \cite{15,16}.

Let's examine structure of the interaction of the $\tau$ lepton to a photon. The most general anomalous vertex function describing $\tau\bar{\tau}\gamma$ interaction for two on-shell $\tau$'s and a photon can be parameterized below \cite{hu,fa},

\begin{eqnarray}
\Gamma^{\nu}=F_{1}(q^{2})\gamma^{\nu}+\frac{i}{2 m_{\tau}}F_{2}(q^{2}) \sigma^{\nu\mu}q_{\mu}+\frac{1}{2 m_{\tau}}F_{3}(q^{2}) \sigma^{\nu\mu}q_{\mu}\gamma^{5}+F_{4}(q^{2})(\gamma^{\nu}-\frac{2m_{\tau}q^{\nu}}{q^{2}})\gamma^{5}
\end{eqnarray}
where $\sigma^{\nu\mu}=\frac{i}{2}(\gamma^{\nu}\gamma^{\mu}-\gamma^{\mu}\gamma^{\nu})$, $q$ is the momentum transfer to the photon and $m_{\tau}=1.777$ GeV is the mass of the $\tau$ lepton. The $q^{2}$-dependent form factors $F_{1}(q^{2}),F_{2}(q^{2})$ and $F_{3}(q^{2})$ have familiar interpretations in limit $q^{2} \rightarrow 0$:

\begin{eqnarray}
F_{1}(0)=1,\: F_{2}(0)=a_{\tau},\: F_{3}(0)=\frac{2m_{\tau}d_{\tau}}{e}.
\end{eqnarray}

In many studies investigating the anomalous magnetic and electric dipole moments of the $\tau$ lepton, photon or $\tau$ leptons in $\tau\bar{\tau}\gamma$ couplings in the examined processes are off-shell. Then, the quantity investigated in these studies is not actually the anomalous $a_{\tau}$ and $d_{\tau}$ couplings due to the $\tau$ lepton is off-shell. For this reason, instead of $a_{\tau}$ and $d_{\tau}$ we can call the anomalous magnetic and electric dipole moments of $\tau$ lepton examined as $\tilde{a}_{\tau}$ and $\tilde{d}_{\tau}$. Therefore, the possible deviation from the SM predictions of $\tau\bar{\tau}\gamma$ couplings could be investigated in a model independent way by means of the effective Lagrangian approach. In this approach, the anomalous $\tau\bar{\tau}\gamma$ couplings are described by means of high-dimensional effective operators. In our analysis, we assume the dimension-six effective operators that contribute to the electromagnetic dipole moments of the $\tau$ lepton.

The experimental bounds on the $\tilde{a}_{\tau}$ at $95\%$ Confidence Level (C.L.) are provided by L3
and OPAL Collaborations through the reaction $e^{-} e^{+}\rightarrow \tau^{-} \tau^{+} \gamma$ at LEP at $\sqrt{s}=M_{Z}$ \cite{4,5}

\begin{eqnarray}
\text{L3}:-0.052<\tilde{a}_{\tau}<0.058,
\end{eqnarray}
\begin{eqnarray}
\text{OPAL}: -0.068<\tilde{a}_{\tau}<0.065.
\end{eqnarray}

The present most restrictive bounds on the $\tilde{a}_{\tau}$ are obtained by the DELPHI Collaboration from the process $e^{-} e^{+}\rightarrow e^{-}  \gamma^{*} \gamma^{*} e^{+} \rightarrow e^{-} \tau^{-} \tau^{+} e^{+}$ total cross section measurements at $\sqrt{s}=183-208$ GeV \cite{6}

\begin{eqnarray}
-0.052<\tilde{a}_{\tau}<0.013
\end{eqnarray}

The present experimental limits on the anomalous $\tilde{d}_{\tau}$ coupling of the $\tau$ lepton at the LEP by L3, OPAL and DELPHI Collaborations are

\begin{eqnarray}
\text{L3}:|\tilde{d}_{\tau}|<3.1 \times 10^{-16}\, e\,cm,
\end{eqnarray}
\begin{eqnarray}
\text{OPAL}:|\tilde{d}_{\tau}|<3.7 \times 10^{-16}\, e\,cm,
\end{eqnarray}
\begin{eqnarray}
\text{DELPHI}:|\tilde{d}_{\tau}|<3.7 \times 10^{-16}\, e\,cm.
\end{eqnarray}

Besides, the most restrictive experimental bounds are given by BELLE Collaboration \cite{bel},

\begin{eqnarray}
-2.2<Re(\tilde{d}_{\tau})<4.5 \times (10^{-17}\, e\,cm),
\end{eqnarray}
\begin{eqnarray}
-2.5<Im(\tilde{d}_{\tau})<0.8 \times (10^{-17}\, e\,cm).
\end{eqnarray}

In the literature, there have been many studies for the anomalous $\tilde{a}_{\tau}$ and $\tilde{d}_{\tau}$ couplings at linear and hadron colliders. The linear colliders and their operating modes of $e \gamma$ and $\gamma \gamma$ have analyzed via the processes $e^{-} e^{+}\rightarrow \tau^{-} \tau^{+} \gamma$ \cite{4},  $e^{-} e^{+}\rightarrow \tau^{-} \tau^{+}$ \cite{ee,ee1}, $e^{-} e^{+}\rightarrow e^{-}  \gamma^{*} \gamma^{*} e^{+} \rightarrow e^{-} \tau^{-} \tau^{+} e^{+}$ \cite{al,sat}, $e^{-} e^{+}\rightarrow e^{-}  \gamma^{*} e^{+} \rightarrow \nu_{e} \tau \bar{\nu}_{\tau} e^{+} $ \cite{al2}, $\gamma \gamma \rightarrow \tau^{+} \tau^{-}$ \cite{al3}, $\gamma \gamma \rightarrow \tau^{+} \gamma\tau^{-}$ \cite{al3} and $Z\rightarrow \tau^{-} \tau^{+} \gamma$ \cite{al4}. Also, these couplings at the LHC have examined through the processes $p p\rightarrow p  \gamma^{*} \gamma^{*} p \rightarrow p \tau^{-} \tau^{+} p$ \cite{al5}, $p p\rightarrow p  \gamma^{*}p \rightarrow p \tau \tau^{+} \bar{\nu}_{\tau} q^{\prime} p$ \cite{al6}, $H \rightarrow \tau^{-} \tau^{+} \gamma$ \cite{al7}. Finally, there is a lot of work related to the $\tilde{a}_{\tau}$ anomalous magnetic moment of the $\tau$ lepton \cite{al8,al9,al10,al11,y1,y2,y3,y4,y5,y6,y7}. All of the experimental and theoretical limits on the electric and magnetic dipole moments of the tau lepton are given in Table I.

In the investigated processes for examining the electromagnetic dipole moments of the $\tau$ lepton in particle accelerators, it is not possible that all the particles are on-shell. For this reason, we can use the effective Lagrangian approach to study the anomalous magnetic and electric dipole moments of this particle. In our analysis, we consider dimension-six operators mentioned in Ref. \cite{al12} related to the electromagnetic dipole moments of the $\tau$ lepton. These operators are given as follows

\begin{eqnarray}
Q_{LW}^{33}=(\bar{\ell_{\tau}}\sigma^{\mu\nu}\tau_{R})\sigma^{I}\varphi W_{\mu\nu}^{I}
\end{eqnarray}

\begin{eqnarray}
Q_{LB}^{33}=(\bar{\ell_{\tau}}\sigma^{\mu\nu}\tau_{R})\varphi B_{\mu\nu}.
\end{eqnarray}

\noindent where $\varphi$ and $\ell_{\tau}$ are the Higgs and the left-handed $SU(2)$ doublets, $\sigma^{I}$ are the Pauli
matrices and  $W_{\mu\nu}^{I}$ and $B_{\mu\nu} $ are the gauge field strength tensors. Therefore, the effective Lagrangian is parameterized as follows,

\begin{eqnarray}
L_{eff}=\frac{1}{\Lambda^{2}} [C_{LW}^{33} Q_{LW}^{33}+C_{LB}^{33} Q_{LB}^{33}+h.c.].
\end{eqnarray}

\noindent After the electroweak symmetry breaking, contributions to the electromagnetic dipole moments of the $\tau$ lepton can be written as

\begin{eqnarray}
\kappa=\frac{2 m_{\tau}}{e} \frac{\sqrt{2}\upsilon}{\Lambda^{2}} Re[\cos\theta _{W} C_{LB}^{33}- \sin\theta _{W} C_{LW}^{33}]
\end{eqnarray}

\begin{eqnarray}
\tilde{\kappa}=-\frac{\sqrt{2}\upsilon}{\Lambda^{2}} Im[\cos\theta _{W} C_{LB}^{33}- \sin\theta _{W} C_{LW}^{33}],
\end{eqnarray}

\noindent where $\upsilon$ is the vacuum expectation value and $\sin\theta _{W}$ is the weak mixing angle.
The relations of CP even parameter $\kappa$ and CP odd parameter $\tilde{\kappa}$ to the anomalous magnetic and electric dipole moments of the $\tau$ lepton are given as follows

\begin{eqnarray}
\kappa=\tilde{a}_\tau, \;\;\;\; \tilde{\kappa}=\frac{2m_\tau}{e}\tilde{d}_\tau.
\end{eqnarray}

The LHC may not provide highly precision measurements due to strong interactions of $pp$ collisions. An $ep$ collider may be a good idea to complement the LHC physics program. Since $ep$ colliders have high center-of-mass energy and high luminosity, new physics effects beyond the SM may appear by examining the interaction of the $\tau$ lepton with photon which requires to measure $\tau \bar{\tau} \gamma$ couplings precisely. The Large Hadron Electron Collider (LHeC) and the Future Circular Hadron Electron Collider (FCC-he) are planned to generate $ep$ collisions at energies from $1.30$ TeV to $10$ TeV \cite{al13,al14}. The LHeC is a suggested deep inelastic electron-nucleon scattering machine which has been planned to collide electrons with an energy from $60$ GeV to possibly $140$ GeV, with protons with an energy of $7$ TeV. In addition, FCC-he is designed electrons with an energy from $250$ GeV to $500$ GeV, with protons with an energy of $50$ TeV.

The remainder of the study is structured as follows: In Section II, we observe the total cross
sections and the anomalous magnetic and electric dipole moments of the $\tau$ lepton via the process $ep \rightarrow e\gamma^{*} \gamma^{*} p\rightarrow e\tau \bar{\tau}p$. Finally, we discuss the conclusions in Section III.

\section{The Total Cross Sections}

The well-known applications of $ep$ colliders are $e\gamma^{*}$, $\gamma^{*}p$ and $\gamma^{*} \gamma^{*}$ collisions where the emitted quasireal photon $\gamma^{*}$ is scattered with small angles from the beam pipe of electron or proton beams. Since $\gamma^{*}$ has a low virtuality, it is almost on the mass shell. $e\gamma^{*}$, $\gamma^{*}p$ and $\gamma^{*} \gamma^{*}$ collisions are defined by the Weizsacker-Williams Approximation (WWA). This approximation has many advantages. It helps to obtain crude numerical estimates through simple formulas. Furthermore, this approach may principally ease the experimental analysis because it gives an opportunity one to directly achieve a rough cross section for $\gamma^{*} \gamma^{*}\rightarrow X$ subprocess through the research of the reaction $ep \rightarrow e X p$ where $X$ symbolizes objects generated in the final state. Nevertheless, in many studies, new physics investigations are examined by using the WWA \cite{s1,s2,s3,s4,s5,s6,s7,s8,s9,s10,s11,s12,s13,s14,s15,s16,s17,s18,s19,s20,s21,s22,sa1,sa2,sa3,sa4,sa5,sa6,sa7,sa8,sa9,sa10,sa11}.

The phenomenological investigations at $ep$ colliders generally contain usual deep inelastic scattering reactions where the colliding proton
dissociates into partons. Although inelastic processes have been more examined in literature, elastic production processes have been less probed. On the other hand, elastic or exclusive processes are occasionally called two photon processes. In two photon processes, photons emitted by electron and proton carry a small amount of virtuality. If photon emitted by proton has high virtuality, proton dissociates after the emission. Thus, photon emitting intact proton deviate slightly from their trajectory along the beam path.

Exclusive processes can be distinguished from completely inelastic processes due to some experimental signatures.  First, after the elastic emission of two photons, electron and proton are scattered with a small angle and escape detection from the central detectors. This gives rise to a missing energy signature called forward large-rapidity gap, in the corresponding forward region of the central detector. However, productions $\ell \bar{\ell}$, $\gamma\gamma$, $jj$ and $J/\psi$ of exclusive processes with the aid of this technique were successfully examined by CDF and CMS Collaborations \cite{fer1,fer2,fer3,cms,cms1}. Also, another experimental signature can be implemented by forward particle tagging. These detectors are to tag the electrons and protons with some energy fraction loss. One of the well known applications of the forward detectors is the high energy photon induced interaction with exclusive two lepton final states. Two quasireal photons emitted by electron and proton beams interact each other to produce two leptons $\gamma^{*} \gamma^{*}\rightarrow \ell^{-} \ell^{+}$.  Deflected electrons and protons and their energy loss will be detected by the forward detectors mentioned above but leptons in the final state will go to the central detector. Produced lepton pairs have very small backgrounds \cite{lep}. Finally, operation of forward detectors in conjunction with central detectors with precise timing, can efficiently reduce backgrounds. CMS and TOTEM Collaborations at the LHC began these measurements using forward detectors between the CMS interaction point and detectors in the TOTEM area about 210 m away on both sides of interaction point \cite{1}. However, LHeC Collaboration has a program of forward physics with extra detectors located in a region between a few tens up to several hundreds of metres from the interaction point \cite{2}.

The $\gamma^{*}$ photons emitted from both electron and proton beams collide with each other, and $\gamma^{*} \gamma^{*}$ collisions are generated. The process $\gamma^{*} \gamma^{*} \rightarrow \tau \bar{\tau}$ participates as a subprocess in the process $ep \rightarrow e\gamma^{*} \gamma^{*} p\rightarrow e\tau \bar{\tau}p$. The Feynman diagrams for the subprocess $\gamma^{*} \gamma^{*} \rightarrow \tau \bar{\tau}$ are shown in Fig. 1. In addition, the diagram of the process $ep \rightarrow e\gamma^{*} \gamma^{*} p\rightarrow e\tau \bar{\tau}p$ is given in Fig. 2.

For the subprocess $\gamma^{*} \gamma^{*} \rightarrow \tau \bar{\tau}$ that has two Feynman diagrams, the polarization
summed amplitude square are obtained as follows

\begin{eqnarray}
|M_{1}|^{2}=&&\frac{16\pi^{2}Q_{\tau}^{2}\alpha^{2}}{2m_{\tau}^{4}(\hat{t}-m_{\tau}^{2})^{2}}[48\kappa(m_{\tau}^{2}-\hat{t}) \nonumber \\
&&(m_{\tau}^{2}+\hat{s}-\hat{t})m_{\tau}^{4}-16(3m_{\tau}^{4}-m_{\tau}^{2}\hat{s}+\hat{t}(\hat{s}+\hat{t})) \nonumber \\
&&m_{\tau}^{4}+2(m_{\tau}^{2}-\hat{t})(\kappa^{2}(17m_{\tau}^{4}+(22\hat{s}-26\hat{t})m_{\tau}^{2}+\hat{t} \nonumber \\
&&(9\hat{t}-4\hat{s}))+\tilde{\kappa}^{2}(17m_{\tau}^{2}+4\hat{s}-9\hat{t})(m_{\tau}^{2}-\hat{t}))m_{\tau}^{2}+12\kappa \nonumber \\
&&(\kappa^{2}+\tilde{\kappa}^{2})\hat{s}(m_{\tau}^{3}-m_{\tau}\hat{t})^{2}-(\kappa^{2}+\tilde{\kappa}^{2})^{2}(m_{\tau}^{2}-\hat{t})^{3} \nonumber \\
&&(m_{\tau}^{2}-\hat{s}-\hat{t}) \nonumber \\
\end{eqnarray}

\begin{eqnarray}
|M_{2}|^{2}=&&\frac{-16\pi^{2}Q_{\tau}^{2}\alpha^{2}}{2m_{\tau}^{4}(\hat{u}-m_{\tau}^{2})^{2}}[48\kappa(m_{\tau}^{4}+(\hat{s}-2\hat{t}) \nonumber \\
&&m_{\tau}^{2}+\hat{t}(\hat{s}+\hat{t}))m_{\tau}^{4}+16(7m_{\tau}^{4}-(3\hat{s}+4\hat{t})m_{\tau}^{2}+\hat{t}(\hat{s}+\hat{t})) \nonumber \\
&&m_{\tau}^{4}+2(m_{\tau}^{2}-\hat{t})(\kappa^{2}(m_{\tau}^{4}+(17\hat{s}-10\hat{t})m_{\tau}^{2}+9\hat{t}(\hat{s}+\hat{t})) \nonumber \\
&&+\tilde{\kappa}^{2}(m_{\tau}^{2}-9\hat{t})(m_{\tau}^{2}-\hat{t}-\hat{s}))m_{\tau}^{2}+(\kappa^{2}+\tilde{\kappa}^{2})^{2}(m_{\tau}^{2}-\hat{t})^{3} \nonumber \\
&&(m_{\tau}^{2}-\hat{s}-\hat{t}) \nonumber \\
\end{eqnarray}

\begin{eqnarray}
|M_{1}^{\dag}M_{2}+M_{2}^{\dag}M_{1}|=&&\frac{16\pi^{2}Q_{\tau}^{2}\alpha^{2}}{m_{\tau}^{2}(\hat{t}-m_{\tau}^{2})(\hat{u}-m_{\tau}^{2})} \nonumber \\
&&[-16(4m_{\tau}^{6}-m_{\tau}^{4}\hat{s})+8\kappa m_{\tau}^{2}(6m_{\tau}^{4}-6m_{\tau}^{2}(\hat{s}+2\hat{t})-\hat{s})^{2} \nonumber \\ &&+6\hat{t})^{2}+6\hat{s}\hat{t})+(\kappa^{2}(16m_{\tau}^{6}-m_{\tau}^{4}(15\hat{s}+32\hat{t})+m_{\tau}^{2}(15\hat{s})^{2} \nonumber \\
&&+14\hat{t}\hat{s}+16\hat{t})^{2})+\hat{s}\hat{t}(\hat{s}+\hat{t}))+\tilde{\kappa}^{2}(16m_{\tau}^{6}-m_{\tau}^{4}(15\hat{s}+32\hat{t}) \nonumber \\
&&+m_{\tau}^{2}(5\hat{s})^{2}+14\hat{t}\hat{s}+16\hat{t})^{2})+\hat{s}\hat{t}(\hat{s}+\hat{t})))-4\kappa(\kappa^{2}+\tilde{\kappa}^{2})\hat{s} \nonumber \\
&&(m_{\tau}^{4}+m_{\tau}^{2}(\hat{s}-2\hat{t})+\hat{t}(\hat{s}+\hat{t}))-2(\kappa^{2}+\tilde{\kappa}^{2})^{2}\hat{s} \nonumber \\
&&(m_{\tau}^{4}-2\hat{t}m_{\tau}^{2}+\hat{t}(\hat{s}+\hat{t}))]. \nonumber \\
\end{eqnarray}
Here, $Q_{\tau}$ shows the $\tau$ lepton charge, $\alpha$ is the fine-structure constant and $\hat{s},\hat{t},\hat{u}$ are the Mandelstam invariants.

In the WWA, two photons are used in the subprocess $\gamma^{*} \gamma^{*} \rightarrow \tau \bar{\tau}$. The spectrum of first photon emitted by electron is given as \cite{we}

\begin{eqnarray}
f_{\gamma^{*}_{1}}(x_{1})=\frac{\alpha}{\pi E_{e}}\{[\frac{1-x_{1}+x_{1}^{2}/2}{x_{1}}]log(\frac{Q_{max}^{2}}{Q_{min}^{2}})-\frac{m_{e}^{2}x_{1}}{Q_{min}^{2}}
&&(1-\frac{Q_{min}^{2}}{Q_{max}^{2}})-\frac{1}{x_{1}}[1-\frac{x_{1}}{2}]^{2}log(\frac{x_{1}^{2}E_{e}^{2}+Q_{max}^{2}}{x_{1}^{2}E_{e}^{2}+Q_{min}^{2}})\} \nonumber \\
\end{eqnarray}
where $x_{1}=E_{\gamma_{1}^{*}}/E_{e}$ and $Q_{max}^{2}$ is maximum virtuality of the photon. Here, we assume $Q_{max}=100$ GeV. The minimum value of $Q_{min}^{2}$ is shown as follows

\begin{eqnarray}
Q_{min}^{2}=\frac{m_{e}^{2}x_{1}^{2}}{1-x_{1}}.
\end{eqnarray}

Second, the spectrum of second photon emitted by proton can be written as follows \cite{we}

\begin{eqnarray}
f_{\gamma^{*}_{2}}(x_{2})=\frac{\alpha}{\pi E_{p}}\{[1-x_{2}][\varphi(\frac{Q_{max}^{2}}{Q_{0}^{2}})-\varphi(\frac{Q_{min}^{2}}{Q_{0}^{2}})]
\end{eqnarray}

where the function $\varphi$ is given by

\begin{eqnarray}
\varphi(\theta)=&&(1+ay)\left[-\textit{In}(1+\frac{1}{\theta})+\sum_{k=1}^{3}\frac{1}{k(1+\theta)^{k}}\right]+\frac{y(1-b)}{4\theta(1+\theta)^{3}} \nonumber \\
&& +c(1+\frac{y}{4})\left[\textit{In}\left(\frac{1-b+\theta}{1+\theta}\right)+\sum_{k=1}^{3}\frac{b^{k}}{k(1+\theta)^{k}}\right]. \nonumber \\
\end{eqnarray}
Here,

\begin{eqnarray}
y=\frac{x_{2}^{2}}{(1-x_{2})},
\end{eqnarray}
\begin{eqnarray}
a=\frac{1+\mu_{p}^{2}}{4}+\frac{4m_{p}^{2}}{Q_{0}^{2}}\approx 7.16,
\end{eqnarray}
\begin{eqnarray}
b=1-\frac{4m_{p}^{2}}{Q_{0}^{2}}\approx -3.96,
\end{eqnarray}
\begin{eqnarray}
c=\frac{\mu_{p}^{2}-1}{b^{4}}\approx 0.028.
\end{eqnarray}
In our calculations, we consider that while the virtuality of the photon emitted by proton is $Q_{max}=1.41$ GeV, the $p_{t}$ cut of outgoing proton is 0.1 GeV.

Therefore, we find the total cross section of the main process $ep \rightarrow e\gamma^{*} \gamma^{*} p\rightarrow e\tau \bar{\tau}p$ by integrating the cross section for the subprocess $\gamma^{*} \gamma^{*} \rightarrow \tau \bar{\tau}$. The total cross section of this process is obtained as follows,

\begin{eqnarray}
\sigma_{ep \rightarrow e\gamma^{*} \gamma^{*} p\rightarrow e\tau \bar{\tau}p}=\int f_{\gamma^{*}_{1}}(x_{1})f_{\gamma^{*}_{2}}(x_{2}) d\hat{\sigma}_{\gamma^{*} \gamma^{*} \rightarrow \tau \bar{\tau}} dx_{1} dx_{2}.
\end{eqnarray}

In this work, all analyzes have been calculated in the CalcHEP package program, including non-standard $\tau\bar{\tau}\gamma$ couplings \cite{al15}. We apply the photon spectrums in the WWA embedded in the CalcHEP. However, tau identification efficiency depends of a specific process, some kinematic parameters and luminosity. Investigations of tau identification have not been examined yet for LHeC and FCC-he detectors. In this case, identification
efficiency can be detected as a function of transverse momentum and rapidity of the tau lepton \cite{tait}. We have considered the following cuts for the selection of tau lepton as used in many studies \cite{as,al5}: $|\eta^{\tau,\bar{\tau}}|<2.5$ and $p_{T}^{\tau,\bar{\tau}}>20$ GeV. These cuts on the tau leptons ensure that their decay products are collimated which allows their momenta to be reconstructed reasonably accurately, despite the unmeasured energy going into neutrinos \cite{tait1}. For this reason, for the process $ep \rightarrow e\gamma^{*} \gamma^{*} p\rightarrow e\tau \bar{\tau}p$, we consider the following basic acceptance cuts to reduce the background and to maximize the signal sensitivity:

\begin{eqnarray}
|\eta^{\tau,\bar{\tau}}|<2.5,
\end{eqnarray}

\begin{eqnarray}
p_{T}^{\tau,\bar{\tau}}>20 \: GeV,
\end{eqnarray}

\begin{eqnarray}
\Delta R_{\tau \bar{\tau}}>0.4.
\end{eqnarray}
Here, $\eta$ is the pseudorapidity which reduces the contamination from other particles misidentified as tau, $p_{T}$ is the transverse momentum cut of the final state particles and $\Delta R$ is the separation of the final state particles.

Using the cuts given above, we give numerical fit functions for the total cross sections of the process $ep \rightarrow e\gamma^{*} \gamma^{*} p\rightarrow e\tau \bar{\tau}p$ as a function of the anomalous couplings for center-of-mass energies of $\sqrt{s}=1.30,1.98,7.07, 10$ TeV. The numerical fit functions for this process are

For $\sqrt{s}=1.30$ TeV
\begin{eqnarray}
\sigma=1.39\times 10^{4}\kappa^{4}+1.62\times 10^{2}\kappa^{3}+1.63\times 10^{2}\kappa^{2}+0.45\kappa+0.135 (pb),
\end{eqnarray}

\begin{eqnarray}
\sigma=1.39\times 10^{4}\tilde{\kappa}^{4}+1.63\times 10^{2}\tilde{\kappa}^{2}+0.135 (pb).
\end{eqnarray}

For $\sqrt{s}=1.98$ TeV
\begin{eqnarray}
\sigma=4.04\times 10^{4}\kappa^{4}+2.83\times 10^{2}\kappa^{3}+2.85\times 10^{2}\kappa^{2}+0.69\kappa+0.211 (pb),
\end{eqnarray}

\begin{eqnarray}
\sigma=4.04\times 10^{4}\tilde{\kappa}^{4}+2.85\times 10^{2}\tilde{\kappa}^{2}+0.211 (pb).
\end{eqnarray}

For $\sqrt{s}=7.07$ TeV
\begin{eqnarray}
\sigma=3.42\times 10^{5}\kappa^{4}+4.73\times 10^{2}\kappa^{3}+4.74\times 10^{2}\kappa^{2}+0.87\kappa+0.272 (pb),
\end{eqnarray}

\begin{eqnarray}
\sigma=3.42\times 10^{5}\tilde{\kappa}^{4}+4.74\times 10^{2}\tilde{\kappa}^{2}+0.272 (pb).
\end{eqnarray}

For $\sqrt{s}=10$ TeV
\begin{eqnarray}
\sigma=9.03\times 10^{5}\kappa^{4}+6.42\times 10^{2}\kappa^{3}+6.43\times 10^{2}\kappa^{2}+1.03\kappa+0.322 (pb),
\end{eqnarray}

\begin{eqnarray}
\sigma=9.03\times 10^{5}\tilde{\kappa}^{4}+6.43\times 10^{2}\tilde{\kappa}^{2}+0.322 (pb).
\end{eqnarray}

In the above equations, the total cross sections at $\kappa=\tilde{\kappa}=0$ give the SM cross section. Also, as can be understood from Eqs. (39)-(46), the linear, quadratic and cubic terms of the anomalous couplings arise from the interference between the SM and anomalous amplitudes, whereas the quartic terms are purely anomalous.

\section{Bounds on the anomalous magnetic and electric dipole moments of the $\tau$ lepton at the LHeC and the FCC-he}

We represent the total cross sections of the process $ep \rightarrow e\gamma^{*} \gamma^{*} p\rightarrow e\tau \bar{\tau}p$ as a function of the anomalous $\kappa$ and $\tilde{\kappa}$ couplings in Figs. $3$-$4$ for center-of-mass energies of $\sqrt{s}=1.30,1.98,7.07,10$ TeV. In this analysis, we consider that only one of the anomalous couplings deviate from the SM at any given time. We can easily understand from these figures that the total cross sections of the examined process increase when the center-of-mass energy increases. In addition, as can be seen from Eqs. 39-46, while the total cross sections are symmetric for the anomalous $\tilde{\kappa}$ coupling, it is nonsymmetric for $\kappa$. For this reason, we expect that while the bounds on the anomalous magnetic dipole moment are asymmetric, the bounds on the electric dipole moment are symmetric. It is easily understood from Figs. $3$-$4$ that the deviation from the SM of the anomalous cross sections containing $\kappa$ and $\tilde{\kappa}$ couplings at $\sqrt{s}=10$ TeV is larger than those of including $\kappa$ and $\tilde{\kappa}$ at $\sqrt{s}=1.30,1.98,7.07$ TeV. Therefore, the obtained bounds on
the anomalous $\kappa$ and $\tilde{\kappa}$ couplings at $\sqrt{s}=10$ TeV are anticipated to be more restrictive than the bounds at $\sqrt{s}=1.30,1.98,7.07$ TeV.

In addition, to visualize the effects of the anomalous $\kappa$ and $\tilde{\kappa}$ couplings on the total cross section of the process $ep \rightarrow e\gamma^{*} \gamma^{*} p\rightarrow e\tau \bar{\tau}p$, we give Figs. 5-8. Figs. 5-8 show that the surfaces of these curves strongly depend on the anomalous $\kappa$ and $\tilde{\kappa}$ couplings.

We need statistical analysis to probe the sensitivity to the anomalous  $\tilde{a}_{\tau}$ and $\tilde{d}_{\tau}$ dipole moments of the $\tau$ lepton. For this reason, we use the usual the $\chi^{2}$ test with a systematic error

\begin{eqnarray}
\chi^{2}=\left(\frac{\sigma_{SM}-\sigma_{NP}}{\sigma_{SM}\delta}\right)^{2},
\end{eqnarray}
where $\sigma_{SM}$ represents only the SM cross section, $\sigma_{NP}$ is the total cross section containing contributions from the SM and new
physics, $\delta=\frac{1}{\sqrt{\delta_{stat}^{2}+\delta_{sys}^{2}}}, \delta_{stat}=\frac{1}{\sqrt{N_{SM}}}$ is the statistical error, $N_{SM}=L_{int}\times BR \times \sigma_{SM}$. The $\tau$ is the only lepton that has the mass necessary to disintegrate, most of the time in hadrons. In $17.8\%$ of the time, the $\tau$ decays into an electron and into two neutrinos; in another $17.4\%$ of the time, it decays in a muon and in two neutrinos. In the remaining $64.8\%$ of the occasions, it decays in the form of hadrons and a neutrino. In our analysis, we assume pure leptonic and semileptonic decays for $\tau^{-}\tau^{+}$ production in the final state of the process. Therefore, we use that branching ratios of the tau pairs are $BR=0.123$ for pure leptonic decays and $BR=0.46$ for semileptonic decays.

Systematic uncertainties may occur in colliders when tau lepton is identified. Due to these uncertainties, tau identification efficiencies are always calculated for specific process, luminosity, and kinematic parameters. These studies are currently being carried out by various groups for selected productions. For a realistic efficiency, we need a detailed study for our specific process and kinematic parameters. On the other hand, in the literature, there are a lot of experimental and theoretical investigations to study the anomalous magnetic and electric dipole moments of the $\tau$ lepton with systematic errors. For example, as seen in Table II, DELPHI Collaboration at the LEP was examined these couplings via the process $e^{-}e^{+}\rightarrow e^{-}\gamma^{*}\gamma^{*}e^{+}\rightarrow e^{-}\tau^{-}\tau^{+}e^{+}$ with systematic errors between $4.3\%$ and $8.9\%$. In Refs. \cite{al5,al6}, the processes $pp\rightarrow p\gamma^{*}\gamma^{*}p\rightarrow p\tau^{-}\tau^{+}p$ and $pp\rightarrow p\gamma^{*}p\rightarrow p\tau \bar{\nu_{\tau}}q^{\prime}$ at the LHC have studied from $2\%$ to $7\%$ with systematic uncertainties. The sensitivity bounds on the anomalous magnetic and electric dipole moments of the $\tau$ lepton through the processes $e^{-}e^{+}\rightarrow e^{-}\gamma^{*}\gamma^{*}e^{+}\rightarrow e^{-}\tau^{-}\tau^{+}e^{+}$, $e^{-}e^{+}\rightarrow e^{-}\gamma^{*}e^{+}\rightarrow \nu_{e}\tau \bar{\nu_{\tau}} e^{+}$, $\gamma \gamma \rightarrow \tau^{-}\tau^{+}$ and $\gamma \gamma \rightarrow \tau^{-}\gamma\tau^{+}$ at the Compact Linear Collider (CLIC) have calculated by considering of systematic errors: $3,5,7\%$ and $10\%$ \cite{al,al12,al13}. On the other hand, we could not have any information on the systematic uncertainties of the process we examined in the LHeC and FCC-he studies. Taking into consideration the previous studies, we consider the total systematic uncertainties of $0,5\%$ and $10\%$.

For pure and semileptonic decay channels, the estimated sensitivities at $95\%$ C.L. on the anomalous  $\tilde{a}_{\tau}$ and $\tilde{d}_{\tau}$ dipole moments of the $\tau$ lepton through the process $ep \rightarrow e\gamma^{*} \gamma^{*} p\rightarrow e\tau \bar{\tau}p$ at the LHeC and FCC-he, as well as for center-of-mass energies of $\sqrt{s}=1.30,1.98,7.07,10$ TeV and systematic errors of $0,5,10\%$ are given in Tables III-X. As can be seen in Table IV, the process $ep \rightarrow e\gamma^{*} \gamma^{*} p\rightarrow e\tau \bar{\tau}p$ at $\sqrt{s}=1.30$ TeV for semileptonic decay channel with $L=100$ fb$^{-1}$ improves approximately the sensitivity of $\tilde{a}_{\tau}$ dipole moment by up to a factor of 8 compared to the LEP. Our bounds on the anomalous $\tilde{d}_{\tau}$ couplings at $\sqrt{s}=1.30$ TeV are competitive with those of LEP. As can be seen in Table X, the best sensitivities obtained on $\tilde{a}_{\tau}$ and $\tilde{d}_{\tau}$ are $-0.0025<\tilde{a}_{\tau}<0.0009$ and $|\tilde{{d}_{\tau}}|<8.85 \times 10^{-18}\, e\,cm$, respectively. Therefore, the collision of FCC-he with $\sqrt{s}=10$ TeV and $L=1000$ fb$^{-1}$ without systematic error probes the anomalous $\tilde{a}_{\tau}$ and $\tilde{d}_{\tau}$ dipole moments with a far better than the experiments bounds. Tables VII-X represent that the bounds with increasing $\delta_{sys}$ values at the FCC-he are almost unchanged with respect to the luminosity values and for the center-of-mass energy values. The reason of this situation is $\delta_{stat}$ which is much smaller than $\delta_{sys}$.

Figs. $9$-$10$ show bounds values obtained the anomalous magnetic and electric dipole moments of the $\tau$ lepton at $95\%$ C.L. through the process $ep \rightarrow e\gamma^{*} \gamma^{*} p\rightarrow e\tau \bar{\tau}p$ at the LHeC and FCC-he for center-of-mass energies of $\sqrt{s}=1.30, 1.98,7.07, 10$ TeV without systematic uncertainty. From these figures we can compare the bounds obtained from four different center-of-mass energies more easily.

We compare our process with photon-photon collisions that are the cleanest process to examine the anomalous $\tilde{a}_{\tau}$ and $\tilde{d}_{\tau}$ dipole moments of the $\tau$ lepton. First, in Ref. \cite{al}, $\gamma^{*} \gamma^{*}$ collisions at the $3$ TeV CLIC with an integrated luminosity of $590$ fb$^{-1}$ show that the bounds on the anomalous $\tilde{a}_{\tau}$ and $\tilde{d}_{\tau}$ couplings are calculated as $-0.0036<a_{\tau}<0.0003$ and $|d_{\tau}|<6.00 \times 10^{-18}\, e\,cm$, respectively. We observe that the bounds obtained from $\gamma^{*} \gamma^{*}$ collisions at the $10$ TeV FCC-he are at the same order with those reported in Ref. \cite{al}.

Another collision studied for investigation the anomalous magnetic and electric dipole moments of the $\tau$ lepton is the process $\gamma \gamma \rightarrow \tau^{-} \tau^{+}$ at the CLIC \cite{al3}. We understand that the sensitivities on $\tilde{a}_{\tau}$ and $\tilde{d}_{\tau}$ dipole moments expected to be obtained for the future $\gamma\gamma$ collisions that generate Compton backscattering photons are roughly 10 times better than our limits.

In addition, we probe the bounds of Ref. \cite{al5}, in which the best bounds on anomalous couplings by examining the the process $pp\rightarrow p\gamma^{*}\gamma^{*}p\rightarrow p\tau^{-}\tau^{+}p$ at the LHC with center-of-mass energy of $14$ TeV and the integrated luminosity of $200$ fb$^{-1}$ are obtained. We see that the anomalous $\tilde{a}_{\tau}$ and $\tilde{d}_{\tau}$ dipole moments we found from our process for semileptonic decay channel with $\sqrt{s}=1.98$ TeV and $L=100$ fb$^{-1}$ are at the same order with those reported in Ref. \cite{al15}. Our best bounds on the anomalous couplings can set more stringent sensitive by one order of magnitude with respect to the best sensitivity derived from $\tau^{+} \tau^{-}$ production at the LHC.

Finally, in Figs. 9-12, we show contours for the anomalous $\kappa$ and $\tilde{\kappa}$ couplings for the process $ep \rightarrow e\gamma^{*} \gamma^{*} p\rightarrow e\tau \bar{\tau}p$ at the LHeC and FCC-he for various integrated luminosities and center-of-mass energies. As we can see from these figures, the improvement in the sensitivity on the anomalous couplings is achieved by increasing to higher center-of-mass energies and luminosities.

\section{Conclusions}

The new physics effects beyond the SM may appear by examining the interaction of the $\tau$ lepton with photon which requires to measure $\tau \bar{\tau} \gamma$ coupling precisely. Nevertheless, the possible deviation from the SM predictions of $\tau \bar{\tau} \gamma$ coupling would be a sign for the presence of new physics beyond the SM. $ep$ colliders with high center-of-mass energy and high luminosity such as the LHeC and FCC-he may be able to provide a lot of
information on new physics beyond the SM as well as providing precise measurements of the SM. For this purpose, possible non-standard $\tau \bar{\tau} \gamma$ coupling at the LHeC and FCC-he is examined in an effective Lagrangian approach. It is usually common to investigate new physics in a model independent way via
effective Lagrangian approach. This approach is defined by high-dimensional operators which lead to anomalous $\tau \bar{\tau} \gamma$ coupling.

In this study, we examine the potential of the process $ep \rightarrow e\gamma^{*} \gamma^{*} p\rightarrow e\tau \bar{\tau}p$ at $ep$ colliders with $\sqrt{s}=1.30, 1.98,7.07, 10$ TeV to study the anomalous dipole moments of the $\tau$ lepton. This process is the cleanest production mechanism for $ep$ colliders. Furthermore, the subprocess $\gamma^{*} \gamma^{*}\rightarrow \tau \bar{\tau}$ isolates $\tau\bar{\tau}\gamma$ coupling which provides the possibility to analyze $\tau\bar{\tau}\gamma$ coupling separately from $\tau\bar{\tau}Z$ coupling. Also, $ep$ colliders with high center-of-mass energy and luminosity are important for new physics research. Since non-standard $\tau \bar{\tau} \gamma$ couplings defined via effective Lagrangian have dimension-six, they have very strong energy dependences. So, the anomalous cross sections including $\tau \bar{\tau} \gamma$ vertex have a higher energy than the SM cross section. Finally, the subprocess $\gamma^{*} \gamma^{*} \rightarrow \tau \bar{\tau}$ may be effective efficient $\tau$ identification due to clean final state when compared to $pp$ collisons of the LHC.

In our analysis, we find that the process $ep \rightarrow e\gamma^{*} \gamma^{*} p\rightarrow e\tau \bar{\tau}p$ at $ep$ colliders lead to a remarkable improvement in the existing experimental bounds on the anomalous $\tilde{a}_{\tau}$ and $\tilde{d}_{\tau}$ couplings. Therefore, we show that $\gamma^{*} \gamma^{*}$ collision at the LHeC and FCC-he are quite suitable for studying the anomalous $\tilde{a}_{\tau}$ and $\tilde{d}_{\tau}$ dipole moments of the $\tau$ lepton.

\pagebreak

\newpage

\begin{figure}
\includegraphics [width=7cm,height=4cm]{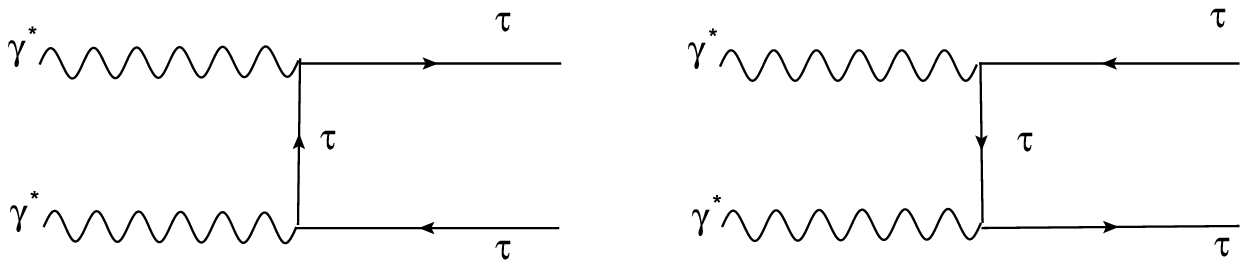}
\caption{Tree-level Feynman diagrams for the subprocess $\gamma^{*} \gamma^{*} \rightarrow \tau \bar{\tau}$.
\label{fig1}}
\end{figure}

\begin{figure}
\includegraphics[width=7cm,height=4cm]{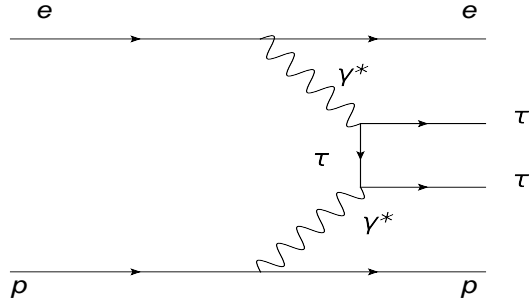}
\caption{Schematic diagram for the process $ep \rightarrow e\gamma^{*} \gamma^{*} p\rightarrow e\tau \bar{\tau}p$.
\label{fig2}}
\end{figure}

\begin{figure}
\includegraphics{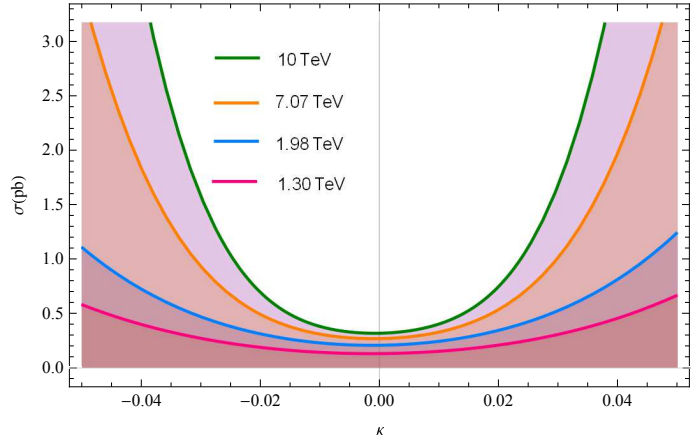}
\caption{The total cross-section of the process $ep \rightarrow e\gamma^{*} \gamma^{*} p\rightarrow e\tau \bar{\tau}p$ as a function of the anomalous $\kappa$ coupling for center-of-mass energies of $\sqrt{s}=1.30,1.98,7.07$ and $10$ TeV.
\label{fig3}}
\end{figure}

\begin{figure}
\includegraphics{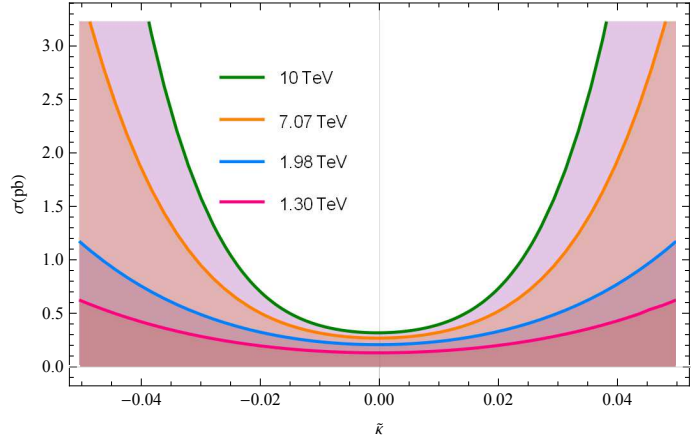}
\caption{Same as in Fig. 3, but for the anomalous $\tilde{\kappa}$ coupling.
\label{fig4}}
\end{figure}

\begin{figure}
\includegraphics{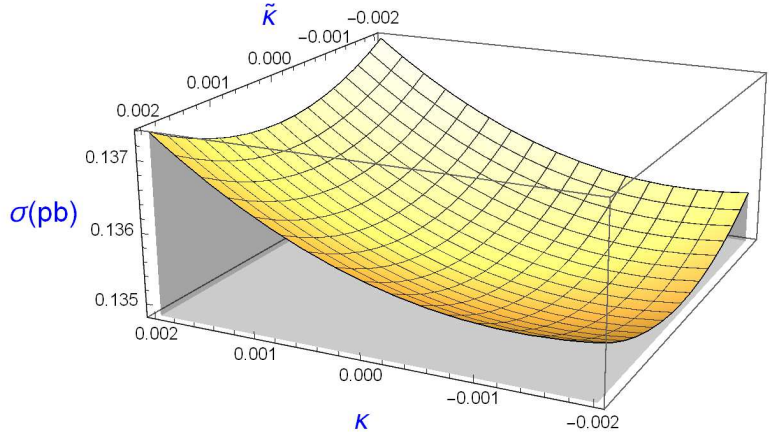}
\caption{The total cross section of the process $ep \rightarrow e\gamma^{*} \gamma^{*} p\rightarrow e\tau \bar{\tau}p$ as a function of the anomalous $\kappa$ and $\tilde{\kappa}$ couplings for center-of-mass energy of $\sqrt{s}=1.30$ TeV.
\label{fig5}}
\end{figure}

\begin{figure}
\includegraphics{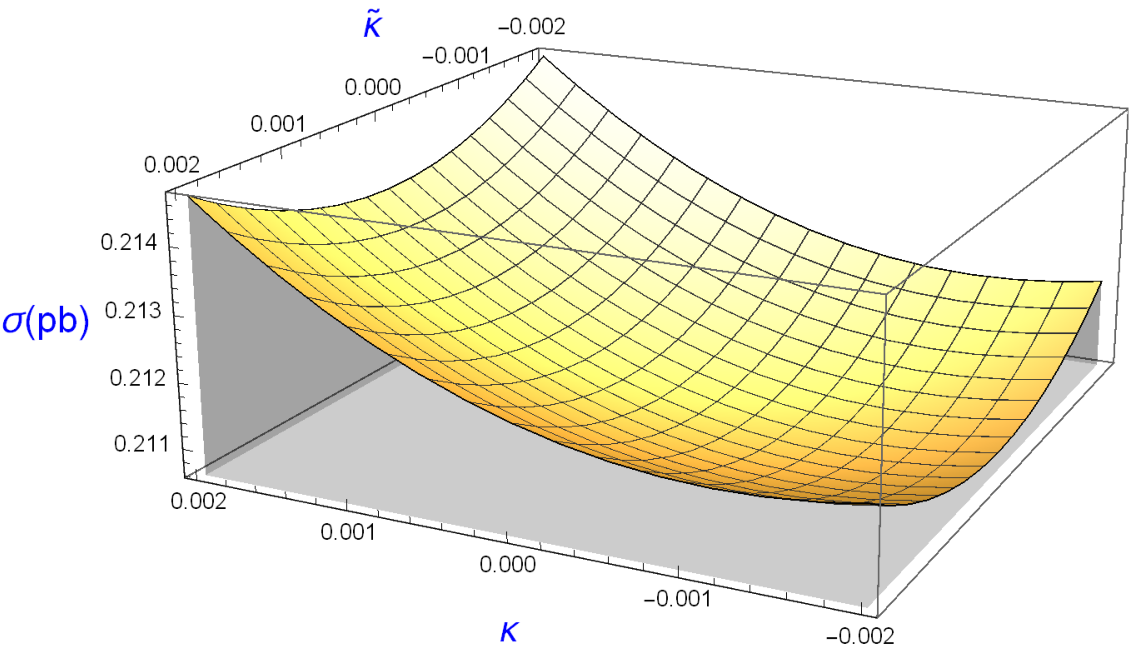}
\caption{Same as in Fig. 5, but for $\sqrt{s}=1.98$ TeV.
\label{fig6}}
\end{figure}

\begin{figure}
\includegraphics{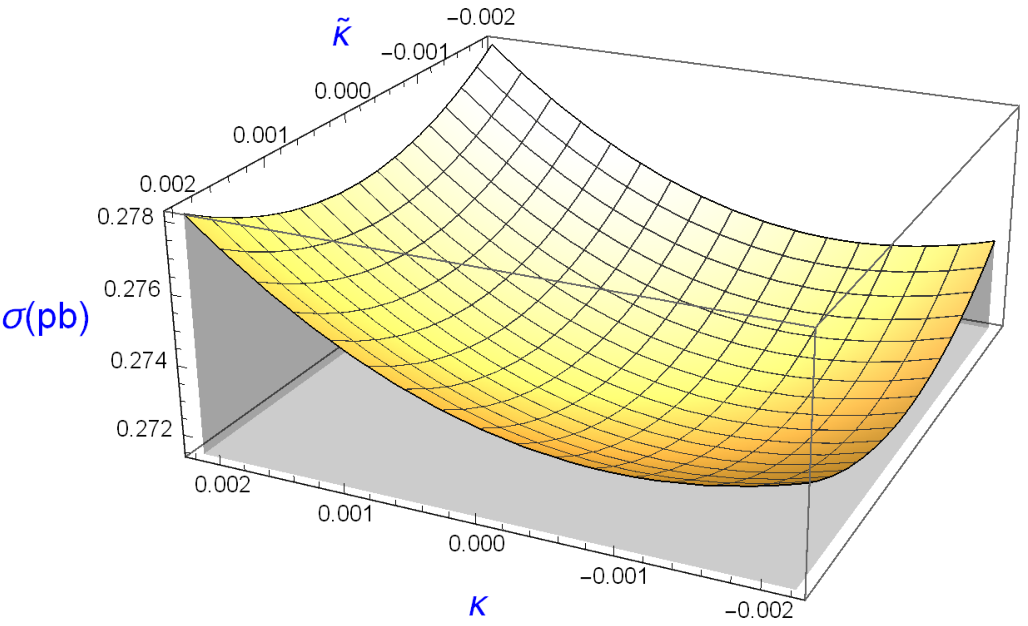}
\caption{Same as in Fig. 5, but for $\sqrt{s}=7.07$ TeV.
\label{fig7}}
\end{figure}

\begin{figure}
\includegraphics{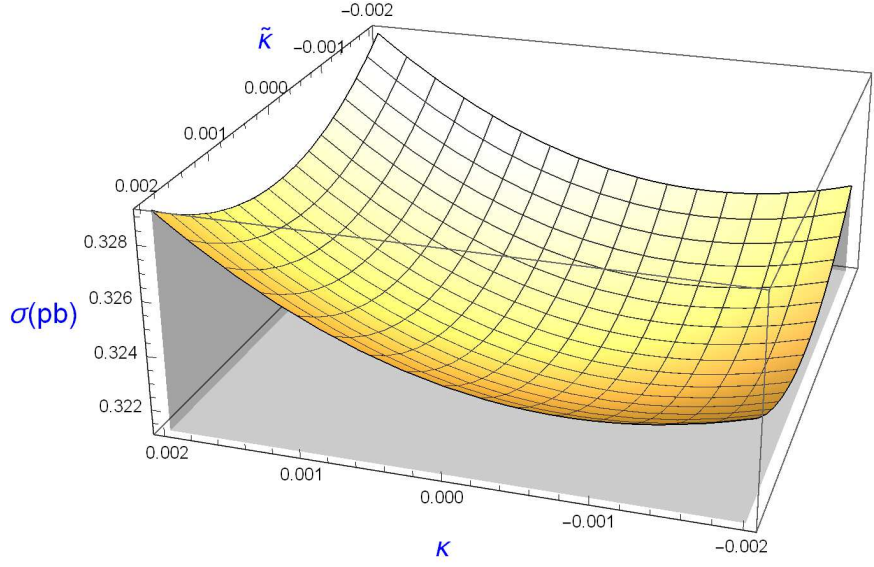}
\caption{Same as in Fig. 5, but for $\sqrt{s}=10$ TeV.
\label{fig8}}
\end{figure}

\begin{figure}
\includegraphics{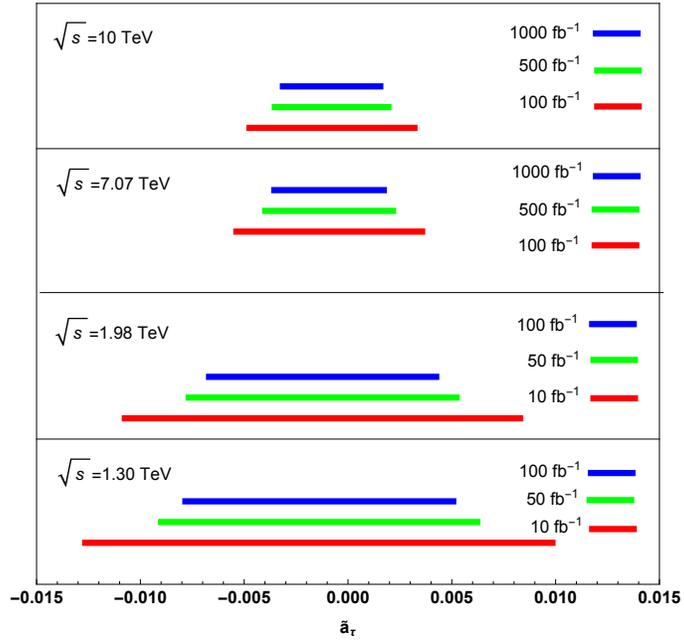}
\caption{Comparison of bounds on $\tilde{a_{\tau}}$ in the process $ep \rightarrow e\gamma^{*} \gamma^{*} p\rightarrow e\tau \bar{\tau}p$ expected at the LHeC and the FCC-he.
\label{fig9}}
\end{figure}

\begin{figure}
\includegraphics{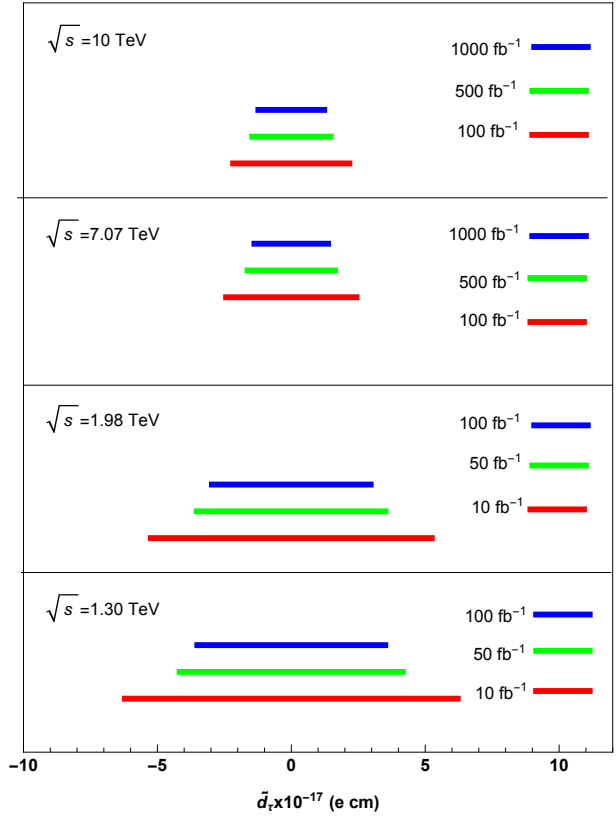}
\caption{Same as in Fig. 9, but for $\tilde{d_{\tau}}$.
\label{fig10}}
\end{figure}

\begin{figure}
\includegraphics{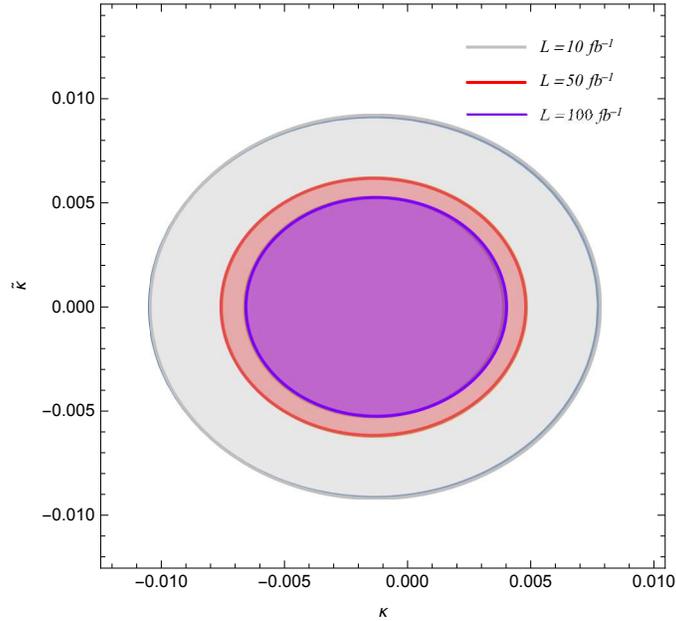}
\caption{Contours for the anomalous $\kappa$ and $\bar{\kappa}$ couplings for the process $ep \rightarrow e\gamma^{*} \gamma^{*} p\rightarrow e\tau \bar{\tau}p$ for center-of-mass energy of $1.30$ TeV.
\label{fig11}}
\end{figure}

\begin{figure}
\includegraphics{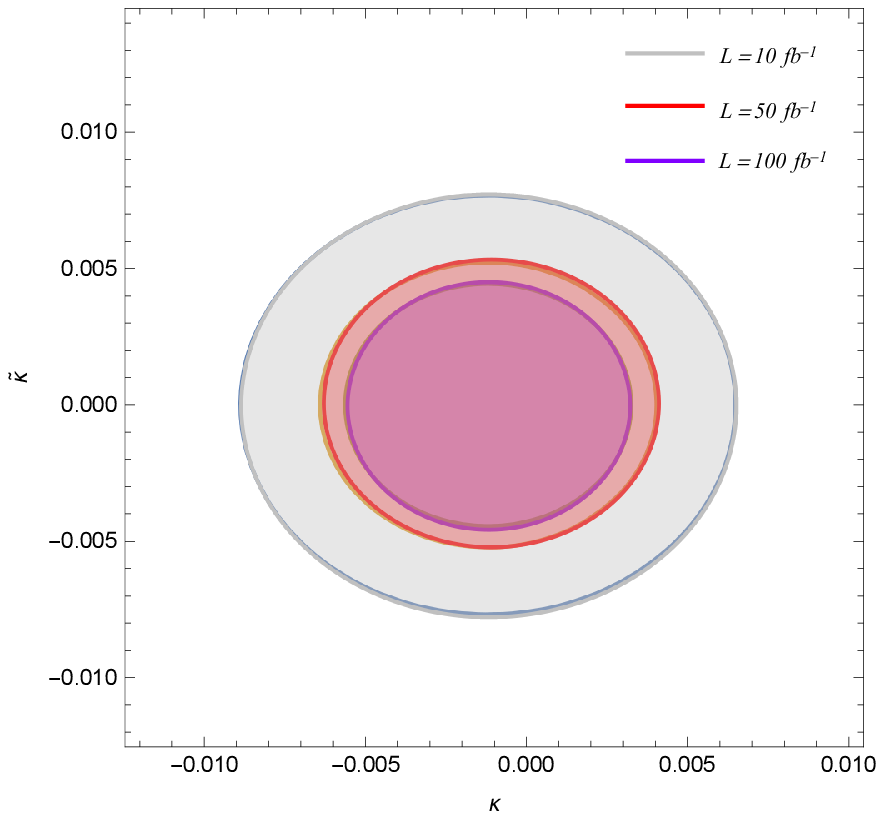}
\caption{Same as in Fig. 11, but for $\sqrt{s}=1.98$ TeV.
\label{fig12}}
\end{figure}

\begin{figure}
\includegraphics{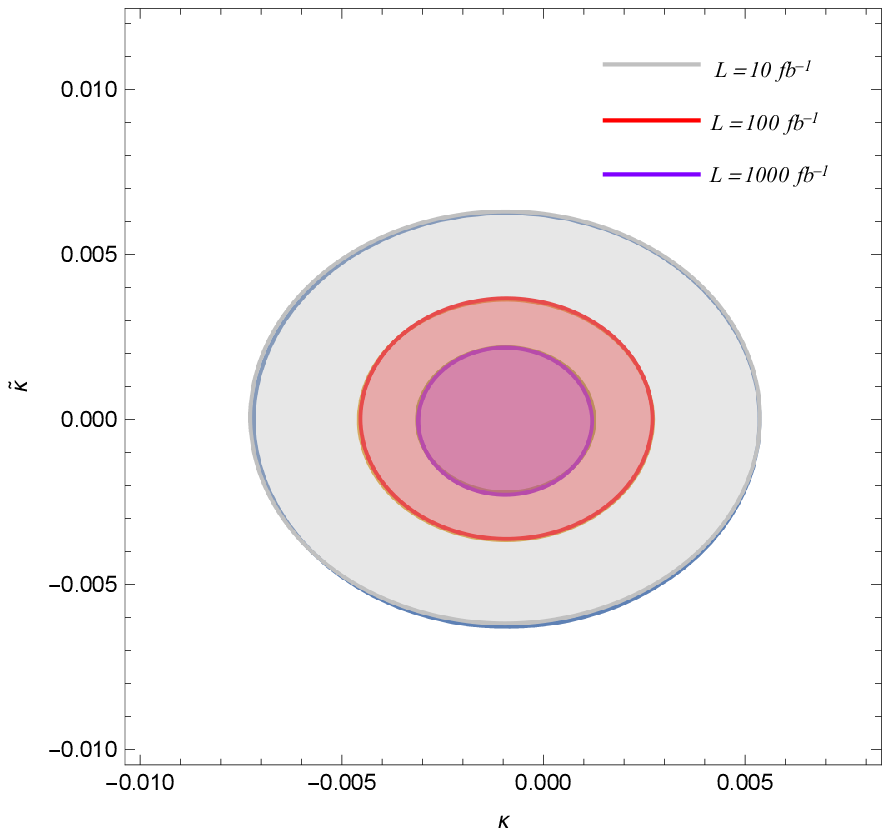}
\caption{Same as in Fig. 11, but for $\sqrt{s}=7.07$ TeV.
\label{fig13}}
\end{figure}

\begin{figure}
\includegraphics{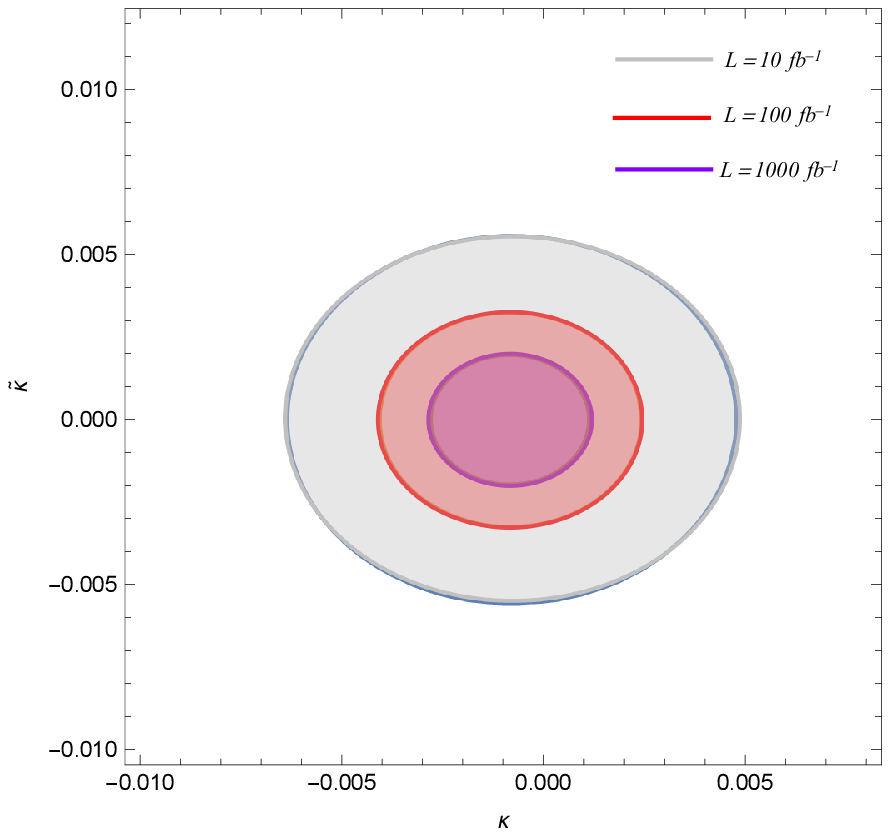}
\caption{Same as in Fig. 11, but for $\sqrt{s}=10$ TeV.
\label{fig14}}
\end{figure}

\begin{table}
\caption{Experimental and theoretical bounds on the electric and magnetic dipole moments of the tau lepton.
\label{tabex}}
\begin{ruledtabular}
\begin{tabular}{cccc}
 & $\tilde{a}_{\tau}$ & $\tilde{d}_{\tau}$ (e\,cm) & Reference \\
\hline
L3& $[-0.052;0.058]$& $<3.1 \times 10^{-16}$& \cite{4} \\
\hline
OPAL& $[-0.068;0.065]$& $<3.7 \times 10^{-16}$& \cite{5} \\
\hline
DELPHI& $[-0.052;0.013]$& $<3.7 \times 10^{-16}$& \cite{6} \\
\hline
BELLE& $-$ & $-2.2<|\textit{Re}|<4.5 (\times 10^{-17})$ \\
& & $-2.5<|\textit{Im}|<0.8 (\times 10^{-17})$ &\cite{bel} \\
\hline
ARGUS& $-$& $|\textit{Re}|<4.6 \times 10^{-16}
|\textit{Im}|<1.8 \times 10^{-16}$& \cite{ee}  \\
\hline
$e^{-} e^{+}\rightarrow e^{-}  \gamma^{*} \gamma^{*} e^{+} \rightarrow e^{-} \tau^{-} \tau^{+} e^{+}$& $[0.0036;0.0003]$& $<6.0 \times 10^{-18}$& \cite{al} \\
\hline
$e^{-} e^{+}\rightarrow e^{-}  \gamma^{*} e^{+} \rightarrow \nu_{e} \tau \bar{\nu}_{\tau} e^{+} $& $[0.0039;0.00038]$& $<2.1 \times 10^{-16}$& \cite{al2} \\
\hline
$\gamma \gamma \rightarrow \tau^{+} \tau^{-}$& $[-0.00015;0.00017]$& $<9.0 \times 10^{-19}$& \cite{al3}  \\
\hline
$\gamma \gamma \rightarrow \tau^{+} \gamma\tau^{-}$& $[-0.00033; 0.00023]$& $<1.5 \times 10^{-18}$& \cite{al3} \\
\hline
$Z\rightarrow \tau^{-} \tau^{+} \gamma$& $-$& $<6.0 \times 10^{-16}$& \cite{al4}  \\
\hline
$p p\rightarrow p  \gamma^{*} \gamma^{*} p \rightarrow p \tau^{-} \tau^{+} p$& $[0.0036;0.00032]$& $<1.4 \times 10^{-17}$& \cite{al5} \\
\hline
$p p\rightarrow p  \gamma^{*}p \rightarrow p \tau \tau^{+} \bar{\nu}_{\tau} q^{\prime} p$& $[0.0037;0.00081]$& $<3.0 \times 10^{-17}$& \cite{al6} \\
\hline
$H \rightarrow \tau^{-} \tau^{+} \gamma$& $[0.0144;0.0106]$& $-$& \cite{al7} \\
\end{tabular}
\end{ruledtabular}
\end{table}

\begin{table}
\caption{Systematic errors given by the DELPHI Collaboration at the LEP.}
\begin{ruledtabular}
\begin{tabular}{ccccc}
 & $1997$& $1998$& $1999$& $2000$ \\
\hline
Trigger efficiency& $7.0$& $2.7$& $3.6$& $4.5$ \\
Selection efficiency& $5.1$& $3.2$& $3.0$& $3.0$  \\
Background& $1.7$& $0.9$& $0.9$& $0.9$  \\
Luminosity& $0.6$& $0.6$& $0.6$& $0.6$ \\
Total& $8.9$& $4.3$& $4.7$& $5.4$  \\
\end{tabular}
\end{ruledtabular}
\end{table}

\begin{table}
\caption{For systematic errors of $0,5\%$ and $10\%$, bounds on the anomalous $\tilde{a}_{\tau}$ and $\tilde{{d}_{\tau}}$ couplings at $\sqrt{s}=1.30$ TeV LHeC via $\tau^{+} \tau^{-}$ production pure leptonic decay channel with integrated luminosities of $10, 30, 50$ and $100$ fb$^{-1}$.}
\begin{ruledtabular}
\begin{tabular}{ccccc}
$\delta_{sys}$& Luminosity($fb^{-1}$)&$\tilde{a}_{\tau}$ & $\vert \tilde{d}_{\tau} \vert (e\,cm) $  \\
\hline
&$10$ &(-$0.0126$, $ 0.0098$)  &$6.2148\times 10^{-17}$ \\
$0\%$&$30$ &(-$0.0100$, $0.0072$) &$ 4.7327\times 10^{-17}$ \\
&$50$ &(-$0.0090$, $0.0062$) &$4.1682\times 10^{-17}$ \\
&$100$ &(-$0.0078$, $0.0050$) &$3.5074\times 10^{-17}$ \\
\hline
&$10$ &(-$0.0136$, $0.0108$) &$6.7716\times 10^{-17}$ \\
$5\%$&$30$ &(-$0.0118$, $ 0.0090$) &$5.7846\times 10^{-17}$ \\
&$50$ &(-$0.0113$, $0.0085$) &$5.5101\times 10^{-17}$ \\
&$100$ &(-$0.0109$, $0.0081$) &$ 5.2729\times 10^{-17}$ \\
\hline
&$10$ &(-$0.0157$, $0.0128$) &$7.9113\times 10^{-17}$ \\
$10\%$& $30$ &(-$0.0147$, $0.0118$) &$7.3692\times 10^{-17}$ \\
&$50$ &(-$0.0145$, $0.0116$) &$7.2445\times 10^{-17}$ \\
&$100$ &(-$0.0143$, $0.0115$) &$  7.1465\times 10^{-17}$ \\
\end{tabular}
\end{ruledtabular}
\end{table}

\begin{table}
\caption{Same as in Table III, but for semileptonic decay channel.}
\begin{ruledtabular}
\begin{tabular}{ccccc}
$\delta_{sys}$& Luminosity($fb^{-1}$)&$\tilde{a}_{\tau}$ & $\vert \tilde{d}_{\tau} \vert (e\,cm) $  \\
\hline
&$10$ &(-$0.0095$, $ 0.0067$)  &$4.4804\times 10^{-17}$ \\
$0\%$&$30$ &(-$0.0076$, $0.0048$) &$ 3.4083\times 10^{-17}$ \\
&$50$ &(-$0.0069$, $0.0041$) &$3.0008\times 10^{-17}$ \\
&$100$ &(-$0.0061$, $0.0033$) &$2.5242\times 10^{-17}$ \\
\hline
&$10$ &(-$0.0116$, $ 0.0088$) &$ 5.6539\times 10^{-17}$ \\
$5\%$&$30$ &(-$0.0109$, $ 0.0081$) &$ 5.2450\times 10^{-17}$ \\
&$50$ &(-$0.0107$, $0.0079$) &$ 5.1504\times 10^{-17}$ \\
&$100$ &(-$0.0106$, $0.0078$) &$5.0759\times 10^{-17}$ \\
\hline
&$10$ &(-$0.0146$, $0.0117$) &$ 7.3083\times 10^{-17}$ \\
$10\%$& $30$ &(-$0.0143$, $ 0.0114$) &$7.1356\times 10^{-17}$ \\
&$50$ &(-$0.0142$, $ 0.0114$) &$7.0995\times 10^{-17}$ \\
&$100$ &(-$0.0141$, $0.0113$) &$  7.1744\times 10^{-17}$ \\
\end{tabular}
\end{ruledtabular}
\end{table}

\begin{table}
\caption{For systematic errors of $0,5\%$ and $10\%$, bounds on the anomalous $\tilde{a}_{\tau}$ and $\tilde{{d}_{\tau}}$ couplings at $\sqrt{s}=1.98$ TeV LHeC via $\tau^{+} \tau^{-}$ production pure leptonic decay channel with integrated luminosities of $10, 30, 50$ and $100$ fb$^{-1}$.}
\begin{ruledtabular}
\begin{tabular}{ccccc}
$\delta_{sys}$& Luminosity($fb^{-1}$)&$\tilde{a}_{\tau}$ & $\vert \tilde{d}_{\tau} \vert (e\,cm) $  \\
\hline
&$10$ &(-$0.0107$, $ 0.0082$)  &$5.2463\times 10^{-17}$ \\
$0\%$&$30$ &(-$0.0085$, $0.0060$) &$ 3.9968\times 10^{-17}$ \\
&$50$ &(-$0.0076$, $0.0052$) &$3.5205\times 10^{-17}$ \\
&$100$ &(-$0.0066$, $0.0042$) &$2.9628\times 10^{-17}$ \\
\hline
&$10$ &(-$0.0119$, $0.0095$) &$ 5.9354\times 10^{-17}$ \\
$5\%$&$30$ &(-$0.0107$, $ 0.0082$) &$5.2243\times 10^{-17}$ \\
&$50$ &(-$0.0103$, $0.0079$) &$5.0396\times 10^{-17}$ \\
&$100$ &(-$0.0101$, $0.0076$) &$ 4.8860\times 10^{-17}$ \\
\hline
&$10$ &(-$0.0142$, $0.0117$) &$7.1869\times 10^{-17}$ \\
$10\%$& $30$ &(-$0.0135$, $0.0111$) &$6.8357\times 10^{-17}$ \\
&$50$ &(-$0.0134$, $0.0109$) &$6.7582\times 10^{-17}$ \\
&$100$ &(-$0.0133$, $ 0.0108$) &$ 6.6982\times 10^{-17}$ \\
\end{tabular}
\end{ruledtabular}
\end{table}

\begin{table}
\caption{Same as in Table V, but for semileptonic decay channel.}
\begin{ruledtabular}
\begin{tabular}{ccccc}
$\delta_{sys}$& Luminosity($fb^{-1}$)&$\tilde{a}_{\tau}$ & $\vert \tilde{d}_{\tau} \vert (e\,cm) $  \\
\hline
&$10$ &(-$0.0081$, $ 0.0056$)  &$3.7839\times 10^{-17}$ \\
$0\%$&$30$ &(-$0.0065$, $0.0041$) &$ 2.8791\times 10^{-17}$ \\
&$50$ &(-$0.0059$, $0.0035$) &$2.5350\times 10^{-17}$ \\
&$100$ &(-$0.0052$, $0.0028$) &$2.1326\times 10^{-17}$ \\
\hline
&$10$ &(-$0.0105$, $ 0.0080$) &$5.1355\times 10^{-17}$ \\
$5\%$&$30$ &(-$0.0100$, $ 0.0076$) &$4.8684\times 10^{-17}$ \\
&$50$ &(-$0.0099$, $ 0.0075$) &$ 4.8092\times 10^{-17}$ \\
&$100$ &(-$0.0098$, $0.0074$) &$ 4.7633\times 10^{-17}$ \\
\hline
&$10$ &(-$0.0135$, $0.0110$) &$6.7977\times 10^{-17}$ \\
$10\%$& $30$ &(-$0.0133$, $0.0108$) &$6.6916\times 10^{-17}$ \\
&$50$ &(-$0.0132$, $ 0.0108$) &$6.6698\times 10^{-17}$ \\
&$100$ &(-$0.0132$, $ 0.0108$) &$ 6.6532\times 10^{-17}$ \\
\end{tabular}
\end{ruledtabular}
\end{table}

\begin{table}
\caption{For systematic errors of $0,5\%$ and $10\%$, bounds on the anomalous $\tilde{a}_{\tau}$ and $\tilde{{d}_{\tau}}$ couplings at $\sqrt{s}=7.07$ TeV FCC-he via $\tau^{+} \tau^{-}$ production pure leptonic decay channel with integrated luminosities of $100, 300, 500$ and $1000$ fb$^{-1}$.}
\begin{ruledtabular}
\begin{tabular}{ccccc}
$\delta_{sys}$& Luminosity($fb^{-1}$)&$\tilde{a}_{\tau}$ & $\vert \tilde{d}_{\tau} \vert (e\,cm) $  \\
\hline
&$100$ &(-$0.0053$, $0.0035$)  &$2.4346\times 10^{-17}$ \\
$0\%$&$300$ &(-$0.0043$, $0.0025$) &$ 1.8552\times 10^{-17}$ \\
&$500$ &(-$0.0039$, $0.0021$) &$1.6343\times 10^{-17}$ \\
&$1000$ &(-$0.0035$, $0.0017$) &$1.3755\times 10^{-17}$ \\
\hline
&$100$ &(-$0.0084$, $0.0067$) &$4.2036\times 10^{-17}$ \\
$5\%$&$300$ &(-$0.0083$, $0.0066$) &$4.1297\times 10^{-17}$ \\
&$500$ &(-$0.0083$, $0.0065$) &$4.1143\times 10^{-17}$ \\
&$1000$ &(-$0.0083$, $ 0.0065$) &$4.1027\times 10^{-17}$ \\
\hline
&$100$ &(-$0.0111$, $0.0095$) &$5.7258\times 10^{-17}$ \\
$10\%$& $300$ &(-$0.0111$, $ 0.0094$) &$5.6993\times 10^{-17}$ \\
&$500$ &(-$0.0111$, $0.0094$) &$5.6946\times 10^{-17}$ \\
&$1000$ &(-$0.0111$, $ 0.0094$) &$  5.6906\times 10^{-17}$ \\
\end{tabular}
\end{ruledtabular}
\end{table}

\begin{table}
\caption{Same as in Table VII, but for semileptonic decay channel.}
\begin{ruledtabular}
\begin{tabular}{ccccc}
$\delta_{sys}$& Luminosity($fb^{-1}$)&$\tilde{a}_{\tau}$ & $\vert \tilde{d}_{\tau} \vert (e\,cm) $  \\
\hline
&$100$ &(-$0.0041$, $ 0.0023$)  &$ 1.7565\times 10^{-17}$ \\
$0\%$&$300$ &(-$0.0034$, $0.0016$) &$ 1.3367\times 10^{-17}$ \\
&$500$ &(-$0.0032$, $0.0013$) &$1.1725\times 10^{-17}$ \\
&$1000$ &(-$0.0029$, $0.0010$) &$9.9019\times 10^{-18}$ \\
\hline
&$100$ &(-$0.0083$, $ 0.0065$) &$4.1221\times 10^{-17}$ \\
$5\%$&$300$ &(-$0.0083$, $0.0065$) &$ 4.1015\times 10^{-17}$ \\
&$500$ &(-$0.0083$, $0.0065$) &$4.0973\times 10^{-17}$ \\
&$1000$ &(-$0.0082$, $0.0065$) &$4.0942\times 10^{-17}$ \\
\hline
&$100$ &(-$0.0111$, $0.0094$) &$5.6972\times 10^{-17}$ \\
$10\%$& $300$ &(-$0.0111$, $0.0094$) &$5.6902\times 10^{-17}$ \\
&$500$ &(-$0.0111$, $0.0094$) &$5.6888\times 10^{-17}$ \\
&$1000$ &(-$0.0111$, $0.0094$) &$5.6877\times 10^{-17}$ \\
\end{tabular}
\end{ruledtabular}
\end{table}

\begin{table}
\caption{For systematic errors of $0,5\%$ and $10\%$, bounds on the anomalous $\tilde{a}_{\tau}$ and $\tilde{{d}_{\tau}}$ couplings at $\sqrt{s}=10$ TeV FCC-he via $\tau^{+} \tau^{-}$ production pure leptonic decay channel with integrated luminosities of $100, 300, 500$ and $1000$ fb$^{-1}$.}
\begin{ruledtabular}
\begin{tabular}{ccccc}
$\delta_{sys}$& Luminosity($fb^{-1}$)&$\tilde{a}_{\tau}$ & $\vert \tilde{d}_{\tau} \vert (e\,cm) $  \\
\hline
&$100$ &(-$0.0047$, $0.0031$)  &$2.1702\times 10^{-17}$ \\
$0\%$&$300$ &(-$0.0038$, $0.0022$) &$ 1.6562\times 10^{-17}$ \\
&$500$ &(-$0.0035$, $0.0019$) &$1.4597\times 10^{-17}$ \\
&$1000$ &(-$0.0031$, $0.0015$) &$1.2291\times 10^{-17}$ \\
\hline
&$100$ &(-$0.0076$, $0.0062$) &$3.8567\times 10^{-17}$ \\
$5\%$&$300$ &(-$0.0075$, $0.0061$) &$3.7998\times 10^{-17}$ \\
&$500$ &(-$0.0075$, $0.0061$) &$3.7881\times 10^{-17}$ \\
&$1000$ &(-$0.0075$, $0.0060$) &$3.7792\times 10^{-17}$ \\
\hline
&$100$ &(-$0.0100$, $0.0087$) &$5.2219\times 10^{-17}$ \\
$10\%$& $300$ &(-$0.0100$, $ 0.0086$) &$5.2024\times 10^{-17}$ \\
&$500$ &(-$0.0100$, $0.0086$) &$5.1985\times 10^{-17}$ \\
&$1000$ &(-$0.0100$, $ 0.0086$) &$  5.1955\times 10^{-17}$ \\
\end{tabular}
\end{ruledtabular}
\end{table}

\begin{table}
\caption{Same as in Table IX, but for semileptonic decay channel.}
\begin{ruledtabular}
\begin{tabular}{ccccc}
$\delta_{sys}$& Luminosity($fb^{-1}$)&$\tilde{a}_{\tau}$ & $\vert \tilde{d}_{\tau} \vert (e\,cm) $  \\
\hline
&$100$ &(-$0.0038$, $ 0.0023$)  &$1.5684\times 10^{-17}$ \\
$0\%$&$300$ &(-$0.0030$, $0.0015$) &$ 1.1945\times 10^{-17}$ \\
&$500$ &(-$0.0028$, $0.0012$) &$1.0521\times 10^{-17}$ \\
&$1000$ &(-$0.0025$, $0.0009$) &$8.8535\times 10^{-18}$ \\
\hline
&$100$ &(-$0.0075$, $ 0.0061$) &$3.7940\times 10^{-17}$ \\
$5\%$&$300$ &(-$0.0075$, $0.0060$) &$ 3.7782\times 10^{-17}$ \\
&$500$ &(-$0.0075$, $0.0060$) &$ 3.7750\times 10^{-17}$ \\
&$1000$ &(-$0.0075$, $0.0060$) &$3.7726\times 10^{-17}$ \\
\hline
&$100$ &(-$0.0100$, $0.0086$) &$5.2005\times 10^{-17}$ \\
$10\%$& $300$ &(-$0.0100$, $0.0086$) &$5.1952\times 10^{-17}$ \\
&$500$ &(-$0.0100$, $0.0086$) &$5.1942\times 10^{-17}$ \\
&$1000$ &(-$0.0100$, $0.0086$) &$5.1934\times 10^{-17}$ \\
\end{tabular}
\end{ruledtabular}
\end{table}

\end{document}